# A Class of Non Invertible Matrices in GF (2) for Practical One Way Hash Algorithm

Artan Berisha
Faculty of Mathematical and Natural Scienes
University of Prishtina, Kosovo

Behar Baxhaku
Faculty of Mathematical and Natural Scienes
University of Prishtina, Kosovo

Artan Alidema
Faculty of Mathematical and Natural Scienes
University of Prishtina, Kosovo

## ABSTRACT
In this paper, we describe non invertible matrix in GF(2) which can be used as multiplication matrix in Hill Cipher technique for one way hash algorithm. The matrices proposed are permutation matrices with exactly one entry 1 in each row and each column and 0 elsewhere. Such matrices represent a permutation of m elements. Since the invention, Hill cipher algorithm was used for symmetric encryption, where the multiplication matrix is the key. The Hill cipher requires the inverse of the matrix to recover the plaintext from cipher text. We propose a class of matrices in GF(2) which are non invertible and easy to generate.

## 1. INTRODUCTION
The need of keeping secrets and communicating in secure led to development of Cryptography. Cryptography is science of writing secret codes and it dates back from Egyptians with hieroglyph, Romans with substitution cipher (Caesar cipher), Arabs with textual analysis which led to the invention of frequency analysis for breaking mono alphabetic substitution cipher, Germans with Enigma and today the modern cryptography [8]. It transforms readable text to scrambled text; this transformation must be done so that it can be reversible. Cryptography intersects disciplines of Mathematics Computer Science and Technical Sciences. The main purpose of Cryptography is securing and enabling communication between two parties and protecting the data, sensitive data or information from outside attacks. In this context cryptography is based on four specific security requirements: authentication, integrity, privacy and non-repudiation. So the role of cryptography is not only data protection but also provide authentication, there are generally three cryptosystems used to achieve this: symmetric algorithms, asymmetric algorithms and hash algorithms. While symmetric and asymmetric cryptosystem is used for enciphering and deciphering the hash function are used for authentication. The original text is called plaintext and enciphered text is called cipher text. Otherwise the process of deciphering is inverse of enciphering, it has as input the enciphered text and gives as output the plaintext (original text). Both cryptosystems depend on a key, and the difference is that at symmetric cryptosystems one key is used for enciphering and deciphering,

$$E_K(P) = C$$

$$D_K(C) = P$$

$E$ enciphering function, $D$ deciphering function and $K$ key. While in asymmetric cryptosystems the key used for enciphering is called private key and key for deciphering is called public key.

$$E_{K_{PRIVATE}}(P) = C$$

$$D_{K_{PUBLIC}}(C) = P$$

The third cryptosystem mentioned is used for authentication, integrity that for any amount of data always gives a fixed output. This output is called hash value.

In this paper we will discuss hash algorithms and give a solution to the one way hash algorithm proposed by [5]. In [5] it is mentioned the need for designing an algorithm for non invertible matrix and then design one way algorithm to generate hash value. Hash algorithm proposed by [5] is based on matrix encryption algorithm called matrix cipher. In the matrix cipher the plaintext is enciphered by transformation [7]

$$C = AP + B \ (mod \ n)$$

the matrix $A$ is $m \times m$ matrix (called enciphering matrix), $P$ is a column vector corresponding to a block of plaintext of length $m$, and $B$ is a column vector of length $m$. (When B is the zero vector, it is called Hill cipher.) To decipher, we must again solve for $P$:

$$C = AP + B \ (mod \ n)$$
$$AP = C - B \ (mod \ n)$$
$$P = IP = A^{-1}(C - B) \ (mod \ n)$$

If the matrix $A$ is non-invertible then the algorithm will meet the conditions to be one way hash algorithm [5]. From [5] it is clear if the matrix in non-invertible then the process of deciphering is impossible. It is proven that it can be applied to any size of data, produces a fixed output, relatively easily to compute and has one-way property [5].

In this paper first we give introduction and some preliminaries about hash functions, permutation matrices. Then we show the algorithm for generating permutation matrices and with that also the non invertible matrix, following with proposed models and mathematical model of algorithms. At the end we show results gained from experiment and finally conclusion and the future work.

## 2. HASH FUNCTIONS
### 2.1 Definition and description
It is hard to design a function that accepts a variable input and give fixed output with non reversible property. These functions are called hash functions and in real world are built on the idea of a compression function. The inputs to the compression function are a message block and the output is hash of all blocks up to that point. That is
$$H_i = F(B_i, H_{i-1})$$
The hash value in *i*th step becomes input for the hash function in *(i+1)*th step [9].

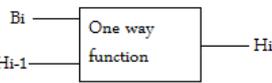

**Fig. 1: One-way hash function scheme**

One-way Hash functions have a important primitive cryptographic, and it is used for authentication, privacy and integrity. The output value from hash function is called hash code [10]. One-way hash function is a function that as a input has variable string length and outputs a fixed binary sequence that cannot be reversed [11]. The usual cryptographic hash functions used today are SHA-1, MD5. As of 2009, the two commonly used algorithms are MD5 and SHA1. Hash function MD5 was broken and it was used to break SSL [9]. Most hash functions are of 64 bits and they are too small to survive a birthday attack that is why most practical hash functions are that produce 128 bit hash code. This will force to hash $2^{64}$ random text to find two hash codes with same value. NIST in its Secure Hash Standard uses 160 bit hash value, this makes even harder for the birthday attack, it requires $2^{80}$ random text to find two hash codes with same value [9]. One-way hash function is an alternative to the message authentication code (MAC), for a variable size input produces a fixed size message [12]. Unlike the MAC, hash code does not require a key as an input to authenticate the message, but a hash value is sent with the message in an authenticated way. The hash value can be encrypted by using symmetric key if the sender and the receiver share the key, or by using public key encryption so the that does not require the keys to be distributed to the parties in communication [14]. Maybe the main role of hash functions is in the provision of digital signatures and authentication. In general hash functions are faster than digital signatures algorithms, the digital signatures are computed to some document by computing the signature on the document's hash value, which is small, compared to the document itself [13]. A hash value can be made public and still not revealing the contents of the document from which is derived. This is important in digital time stamping, because one can get a document time stamped without knowing its contents to the time stamping service [15].

## 2.2 Requirements for one way hash function

For a hash function to be useful for authentication it is necessary to meet these basic requirements are:[14]

- It can applied to any block size of data
- Produces a fixed length output
- Relatively easily to compute (both hardware and software)
- For a given *h* hash value it is hard to find a *x* such *H(x)=h*, this is called one-way property
- For any block size *x*, it is very hard to find $y \neq x$ such *H(y)=H(x)*, this is referred as weak collision resistance
- It is very hard to find pair *(x,y)* such that *H(x)=H(y)*, this is referred as strong collision resistance

The first three properties are required for a practical application of a hash function to message authentication. The fourth states that for a given message it is easy to generate code, but it is impossible to generate a message for a given code. The fifth property states that an alternative message hashing to the same value as a given message cannot be found, and sixth property refers to how resistant the hash function is to a type of attack known as birthday attack [14]. Our proposed model will meet the first three requirements as practical hash function to message authentication.

## 3. PERMUTATION MATRICES DEFINITION AND DESCRIPTION

A square matrix is called permutation matrix if each row and column of the matrix has exactly one 1 and all other entries are 0 [1]. By definition
$$\pi: \{1,2,\ldots,m\} \to \{1,2,\ldots,m\}$$
or
$$\pi = \begin{pmatrix} 1 & 2 & \cdots & m \\ \pi(1) & \pi(2) & \cdots & \pi(m) \end{pmatrix}$$

where $\pi(1), \pi(2), \ldots, \pi(m) \in \{1,2,\ldots,m\}$.
The permutation above can be written as matrix (permutation matrix):
$$P_{ij} = \begin{cases} 1, & if\ i = \pi(j) \\ 0, & otherwise \end{cases}$$

The number of permutation matrices of size n is n! [1]. Permutation matrices are a class of invertible matrices in *GF(2)*. In *GF(2)* each element is either 0 or 1, addition is the binary *exclusive-or* operator (denoted $\oplus$) and multiplication is the binary *and* operator. The arithmetic of row and column in permutation matrix is performed over the commutative ring $\mathbb{Z}/2\mathbb{Z}$ [16]. Permutation matrices have properties that determinant is 1 or -1 and the inverse is the transposed of the matrix. Furthermore the product of two permutation matrices is a permutation matrix.

1) $\det(P) = 1\ or - 1$
2) $PP^T = P^TP = I$, where *I* is identity matrix

When a permutation matrix *P* is multiplied with a matrix *M* from the left it will permute the rows of *M* (the elements of column vector), when *P* is multiplied with *M* from the right it will permute the columns of *M* (the elements of a row vector) [2]. As we can see the permutation matrices are invertible (the property above), but the following lemma will give a result for some property of these matrices [16]. When we refer to a matrix $P^w$, that means that $P^w$, is a square matrix in *GF*(2) with *w* rows and columns.

**Lemma 1**: If some matrix $P^w$, has precisely *w* ones, then $P^w$ is invertible iff it is a permutation matrix [16].

**Lemma 2**: Let $P_1^w$ and $P_2^w$ be permutation matrices. The sum $P_1^w + P_2^w$ is not invertible [16].

**Proof**: Let *P* be the sum of two permutation matrices $P=P_1^w + P_2^w$. We have two cases:

*I*: Suposse there exists *i* and *j* such that $P_1^w(i,j)= P_2^w(i,j) = 1$, then the *i*th row of *P* contains all zeros, so *P* is not invertible.

*II*: Assume that there are no such *i* and *j* as in first case, then *P* will have precisely two ones in each row and column. By induction it is proven that such matrices are not invertible [16].

## 4. GENERATING PERMUTATION MATRICES

After a mathematical background on permutation matrices and their properties we can generate permutation matrices and compute the invertible matrix as sum of two permutation matrices. It is easy to write a permutation matrix following the rule that each row and column of the matrix has exactly one 1 and all other entries are 0. But doing these in random and to gain a permutation matrix it is hard. The first solution will be generating random numbers from *1* to *m* (*m* is the size of matrix) without duplicates. Let *i* be the generated random number, the random number will indicate that the value 1 is at position *(i,1)*, second random number at *(i,2)* and so on till *i* passes throw all values from *1* to *m* with no duplicates. But the generated random numbers are not guaranteed that will generate all numbers in finite time. It could go forever. This algorithm is implemented in some applications, it generates the number from *1* to *m* with duplicates using linear congruential generator with uniform distribution. The better idea is to write the numbers from *1* to *m* and by random to do the permutation. The best algorithm to generate at random a permuted sequence of numbers from *1* to *m* is Fisher-Yates shuffle algorithm or also called Knuth shuffle algorithm [3][4]. The basic of this algorithm is generating a random permutation of numbers from 1 to *m*, and it goes as follows [5]

1. Write down the numbers from 1 to *m*
2. Pick a random number *k* between one and the number of unstruck numbers remaining (inclusive)
3. Counting from the low end, strike out the *k*th number not yet struck out, and write it down elsewhere.
4. Repeat from step 2 until all numbers have been struck out.
5. The sequence of numbers written down in step 3 is now a random permutation of the original numbers.

The code implemented in Java [6]

```
public static void shuffleArray(int[] a) {
    int m = a.length;
    Random random = new Random();
    random.nextInt();
    for (int i = 0; i < m; i++) {
        int j = i + random.nextInt(m - i);
        swap(a, i, j);
    }
}
private static void swap(int[] a, int i, int j) {
    int temp = a[i];
    a[i] = a[j];
    a[j] = temp;
}
```

After generating permuted array $a = a_1 a_2 \ldots a_m$ with random generator, we can create permutation matrices, by setting the value 1 at positions $P(a(1), 1) = P(a(2), 2) = \ldots = P(a(m), m) = 1$.

```
int j = 0;
for (int i = 0; i < m; i ++) {
    P₁[a[i]-1][j] = 1;
    P₂[b[i]-1][j] = 1;
    j = j + 1;
}
```

where *m* is the size of matrix. From the gained matrices we create a sum of them, which is non-invertible matrix as proved above. The implemented code in Java:

```
for (int i = 0; i < m; i ++)
    for (int j = 0; j < m; j++)
        P[i,j] = (P₁[i,j] + P₂[i,j])%2;
```

In the Figure 2 are steps for generating non invertible matrix.

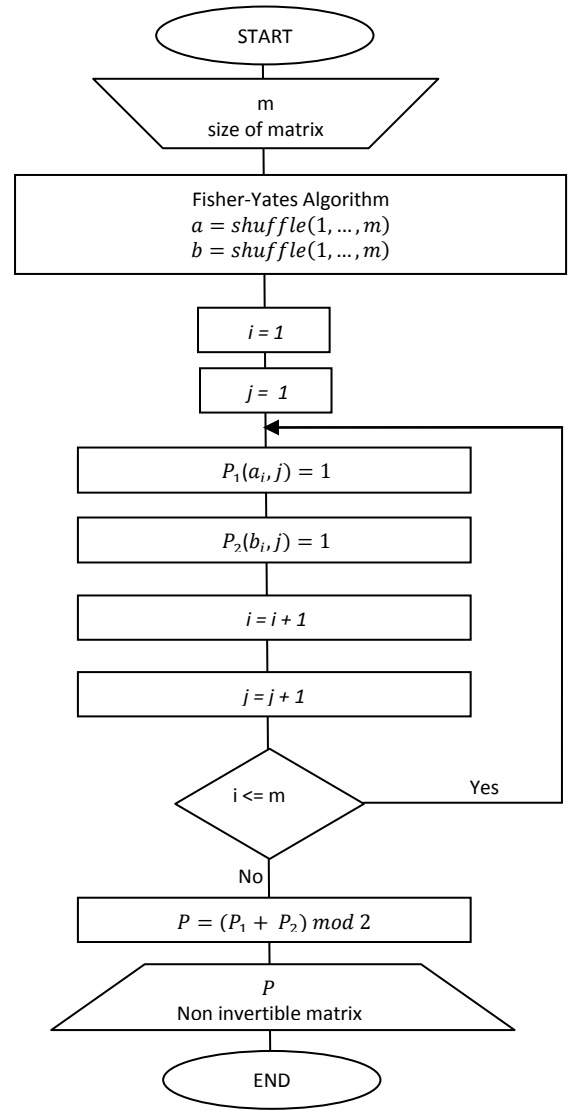

**Fig. 2: Algorithm for generating non invertible matrix in GF(2)**

## 5. PROPOSED ALGORITHM MODELS

In [5] authors proposed a model based on matrix multiplication, it is called Hill cipher, this model is based on the non invertible matrix for practical one-way hash function. This non invertible matrix multiplied plaintext to generate the hash value. We are proposing a solution which automates the model for one-way hash function given by [5]. The non invertible matrix for multiplying plaintext will be generated by given size *m* of the square matrix. Algorithm will generate a non invertible matrix as a sum of two permutation matrices. The elements of the generated matrix will be from GF(2) which means that their value is {0,1}. We are proposing two designs of algorithms for generating hash value. Both models are based in Cipher Block Chaining (CBC), the second model differs from the first in some additional operations. These additional operation steps are to create diffusion using non linear function. Before calculating the hash value the plaintext must be converted to binary data, and divided to column vectors $\{B_1, B_2, ..., B_N\}$ of size *(mx1)*, $B_i = (b_{i0}, b_{i1}, ..., b_{i,m-1})$, $b_{ij} \in \{0,1\}$. Below we show a protocol for using this one way hash function by communication parties.

*Protocol*
Sender: Value $M$ and non invertible matrix $P$
  Calculate $H = A(P,M)$
  Send $H$ to receiver
Receiver: Value $M$ and non invertible matrix $P$
  Calculate $H' = A(P,M)$
  Match $H$ with $H'$

$A(P,M)$ is algorithm proposed below.

*Algorithm of proposed model*
INPUT:  Non invertible matrix $P$ of size $m$ and value $M$.
OUTPUT:  $m$ bit hash value.
STEP 1  Convert value $M$ to binary form $b_0, b_1, ..., b_k$
STEP 2  Padding value algorithm
STEP 3  Initialize $S = 0, H = H^0$,
  $H^0 = h_0^0 h_1^0 ... h_{m-1}^0$, $h_i^0 = 0$ for $i=0,1,...,m-1$.
STEP 4  for $j$ from $0$ to $n-1$
  STEP 5  for $i$ from $0$ to $m-1$
    $M'_{ij} = b_{i+mj}$
STEP 6  for $i$ from $0$ to $n/m-1$
  STEP 6'  if ($i$ mod $2 == 0$ && $i <> 0$)
    $H = F(H, H^i)$
  else for $r$ from $0$ to $m-1$
    $h_r^i = h_r$
  STEP 7  for $j$ from $0$ to $m-1$
    $B_j = M'_{ij} + h_j$
    STEP 8 for $k$ from $0$ to $m-1$
      STEP 9  for $t$ from $0$ to $m-1$
        $S = (P_{kt} * B_t + S)(mod\ 2)$
      $h_k = S$
      $S = 0$

OUTPUT $h_0, h_1, ..., h_{m-1}$.

This is a proposed algorithm for calculating hash value, the step 6' is additional operations for second proposed model (as you can see in Fig. 3). Following we will give definition of non linear function $F(H,H')$ used in second proposed model. The function has two input parameters with m bit each and gives m bit output.

*Function definition*
INPUT:  $H = h_0 h_1 ... h_{m-1}$, $H' = h'_0 h'_1 ... h'_{m-1}$
OUTPUT:  $H = h_0 h_1 ... h_{m-1}$
STEP 1 for $i$ from $0$ to $\left\lfloor\frac{m}{4}\right\rfloor$
$$h_i = \left(h_i + h'_{\left\lfloor\frac{m}{4}\right\rfloor}\right) (mod\ 2)$$
STEP 2 for $i$ from $\left\lfloor\frac{m}{4}+1\right\rfloor$ to $\left\lfloor\frac{m}{2}\right\rfloor$
$$h_i = \left(h_i + h'_{\left\lfloor\frac{m}{4}+1+i\right\rfloor}\right) (mod\ 2)$$
STEP 3 for $i$ from $\left\lfloor\frac{m}{2}+1\right\rfloor$ to $\left\lfloor\frac{3m}{4}\right\rfloor$
$$h_i = \left(h_i + h'_{i-\left\lfloor\frac{m}{2}+1\right\rfloor}\right) (mod\ 2)$$
STEP 3 for $i$ from $\left\lfloor\frac{3m}{4}+1\right\rfloor$ to $m-1$
$$h_i = (h_i + h'_i) (mod\ 2)$$

OUTPUT $H = h_0 h_1 ... h_{m-1}$

Each m bit input will be divide in four blocks, the first block has the bits from 0 to $\left\lfloor\frac{m}{4}\right\rfloor$, second from $\left\lfloor\frac{m}{4}+1\right\rfloor$ to $\left\lfloor\frac{m}{2}\right\rfloor$, third from $\left\lfloor\frac{m}{2}+1\right\rfloor$ to $\left\lfloor\frac{3m}{4}\right\rfloor$ and last one from $\left\lfloor\frac{3m}{4}+1\right\rfloor$ to m-1. The output will be a m bit value as shown in figure 4.

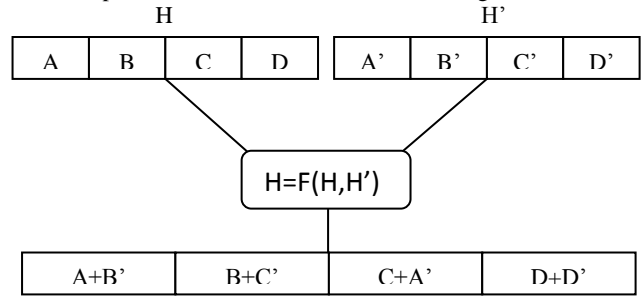

Fig. 4. Definition of F function

The padding for both proposed models is done with same padding algorithm. The padding will be done such that the number *N* of column vectors *(mx1)* will meet the condition N mod 2 =0. There are two cases, if *((k-k mod m)/m) mod 2 =0* then the *(k mod m)* last values will be summed *(mod 2)* with *(k mod m)* first values and the length of output binary values will be *(k-k mod m)*. Second case if the first condition doesn't meet then *(m - k mod m)* values of *0* will be added to input binary sequence. The output length will be *(m+k-k mod m)*.

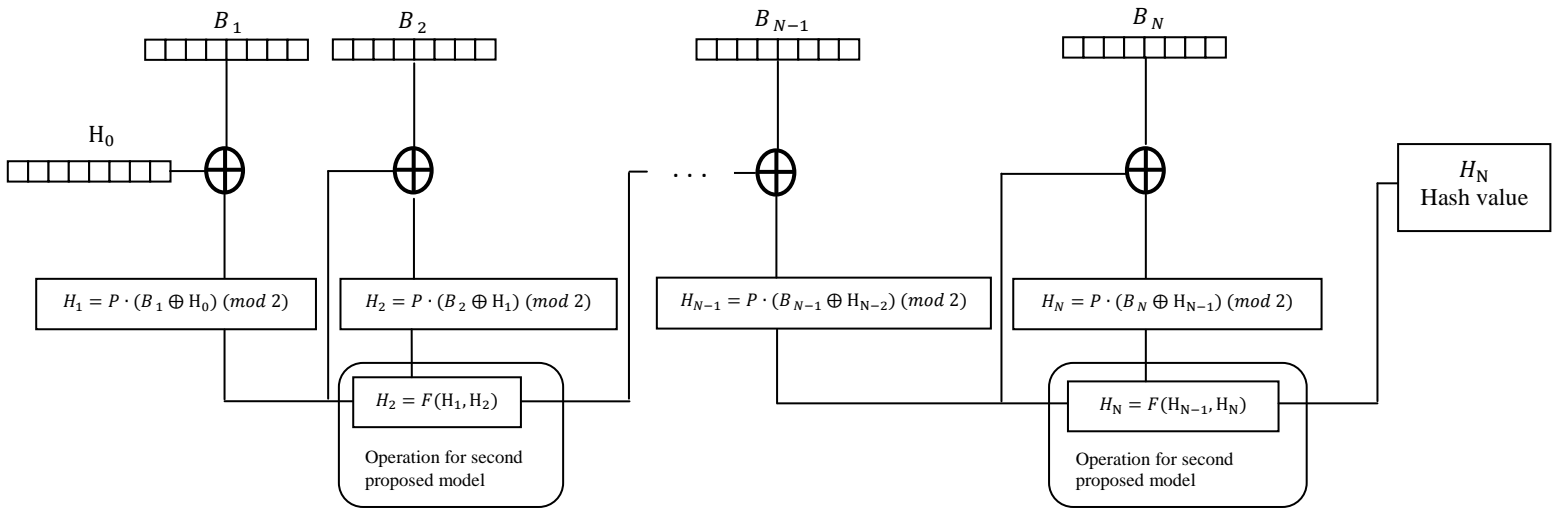

Fig. 3. Proposed one-way hash algorithm model

*Padding Algorithm*
INPUT    $m$, binary values $b_0, b_1, \ldots, b_k$
OUTPUT Binary value $b_0, b_1, \ldots, b_n$   $(n \geq k)$
STEP 1 *if (((k-k mod m)/m) mod 2 ==0)*
          *for i from (k-k mod m+1) to k*
             $b_{i-k+k \bmod m-1} = b_{i-k+k \bmod m-1} + b_i$   $(mod\ 2)$
          else *for i from (k+1) to m+k-k mod m*
             $b_i = 0$
OUTPUT  $b_0, b_1, \ldots, b_k$

## 5.1 Mathematical model of hash algorithm

The mathematical model is based on Hill cipher, but the arithmetic is performed over the commutative ring $\mathbb{Z}/2\mathbb{Z}$. The mathematical description of first proposed model is:

$H_i = P \cdot (B_i \oplus H_{i-1}) \ (mod\ 2), i = 1, 2, \ldots, N$

$N$-number of column vectors, $H_N$–hash value *(mx1)*, $P$- non invertible matrix of size *(mxm)*, $B_i$ –$i$th column vector of plaintext (message) of size *(mx1)*, $H_0$- zero column vector, $\oplus$ - *exclusive-or* operator.

In general the hash value will be (arithmetic used is modulo 2):

$H_N = P \cdot (B_N \oplus P \cdot B_{N-1} \ \oplus \ldots \oplus P^{N-1} \cdot B_1 \oplus P^{N-1} \cdot H_0)$

The mathematical description for second proposed model is:
$H_1 = P \cdot (B_1 \oplus H_0) \ (mod\ 2)$
$H_2 = P \cdot (B_2 \oplus H_1) \ (mod\ 2)$
$H_2 = F(H_1, H_2)$

$H_3 = P \cdot (B_3 \oplus H_2) \ (mod\ 2)$
$H_4 = P \cdot (B_4 \oplus H_3) \ (mod\ 2)$
$H_4 = F(H_3, H_4)$
  .
  .
  .
$H_{N-1} = P \cdot (B_{N-1} \oplus H_{N-2}) \ (mod\ 2)$
$H_N = P \cdot (B_N \oplus H_{N-1}) \ (mod\ 2)$
$H_N = F(H_{N-1}, H_N)$

$H_N -$ is the hash value, F - is a non linear function.

Function F has two parameters, two column vectors with size *(mx1)* and outputs a vector column with length *(mx1)*.

The matrix **P** is non invertible, thus the algorithm is not reversible. In their paper [17] proved that if a non invertible matrix is used for encryption then decryption is not reversible. The number of generated non invertible matrices with size *m* is **m**!, which means that if the size of non invertible matrix is **128** then the space of possible number of non invertible matrices will be $\mathbf{128! \approx 3.85 \cdot 10^{215}}$.

## 5.2 Proof of one-way property for hash algorithm

### 5.2.1 Applied to any size of data
The algorithm works with binary data, so first the plaintext is converted to binary data and then divided to vector columns with size *(mx1)* where *m* is the size of non invertible matrix. This means that can be applied to any amount of data because we can create a matrix with *m* rows and *r* columns.

### 5.2.2 Fixed length output
The algorithm takes a input of variable length and produces a value with fixed length, the length of the hash value is *m*. The number of rounds in one way hash algorithm is $\left\lfloor \frac{N}{m} \right\rfloor$ ($\lfloor x \rfloor$ - floor function, $N$ – size of plaintext in bits, $m$ size of non invertible matrix).

### 5.2.3 Easy to compute
The property *Easy to compute* it is clear from mathematical model. The algorithm needs as input plaintext and size of non invertible matrix. Calculation is done within multiplication of matrix with vector and the *exclusive-or* operator.

### 5.2.4 One-way property
The algorithm model is design based on matrix multiplication, and it is proved from [17] non reversibility of matrix multiplication based cryptosystems.

## 6. COMPARATIVE ANALYSIS

Comparative analyses are done against our proposed algorithm and SHA-2 depending on the time of calculating hash value. We implemented our proposed algorithm and SHA-2 in *Java*. For our experiment we used a test bed with 2.8 GHz and 3 GB RAM. The effect of changing file size for calculating hash value was chosen. Time for calculating hash value was measured from step 6 to step 9 at our proposed algorithm. Same we did for SHA-2 measuring time only in processes that calculated hash value. Below is table with our gained results after the experiment (Table 1.), as we can see the proposed algorithm is faster for the file size less than 6.2 KB, but for large file size it took too much time to calculate hash value (Table 2.).

**Table 1. Time (ms) for calculating hash value with proposed model and SHA-2 for different file size**

| File size (bit) | Calculating hash value (ms)- proposed model | Calculating hash value (ms)- SHA-2 |
|---|---|---|
| 256 | 11 | 36 |
| 512 | 11 | 36 |
| 1024 | 11 | 41 |
| 2048 | 11 | 42 |
| 4096 | 11 | 43 |
| 8192 | 11 | 45 |
| 16384 | 17 | 48 |
| 32768 | 30 | 47 |
| 65536 | 62 | 52 |
| 131072 | 124 | 68 |
| 262144 | 237 | 73 |
| 524288 | 501 | 85 |

The time for calculating hash value at proposed model is increasing very fast after the file size of 32 KB as we can see in Fig. 5.

**Table 2. Extended results of calculating hash value by proposed model for larger file size**

| File size (MB) | 8 | 16 | 32 | 64 | 128 | 256 |
|---|---|---|---|---|---|---|
| Time (min) | 1 | 2 | 3 | 8 | 16 | 32 |

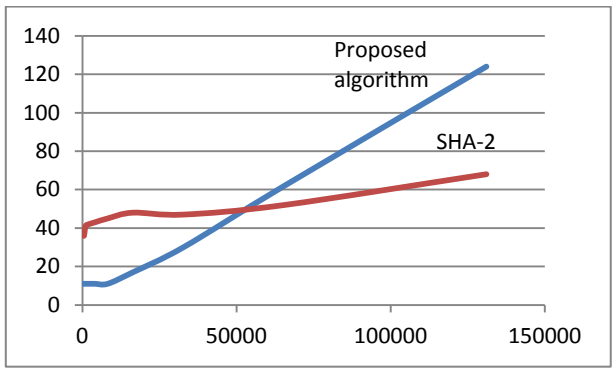

**Fig 5. Graphical view of performance of proposed algorithm against SHA-2**

## 7. CONCLUSION AND FUTURE WORK

In this paper we proposed a technique for generating non invertible matrices in GF(2) and one-way hash algorithm to generate hash value. The non invertible matrices are generated only by giving a size of square matrix. For a small amount of data this proposed model meets all six requirements for one-way hash function. The model proposed is practical for small amount of data, which implies that can be used for authentication (password verification), MAC. For file size less than 6.2 KB it is faster than SHA-2 but for larger amount of data the time to calculate hash value increases exponentially. In future we need to see for non invertible matrices in GF($p^m$) and to speed up the calculating process of hash value.

## 8. ACKNOWLEDGMENT

All the Praises and Thanks be to Allah, the Lord of Heavens and the Lord of the Earth and the Lord of the Worlds.